\documentclass[onecollarge,natbib]{svjour2}
\bibpunct{[}{]}{,}{n}{}{,} 
\smartqed  
\usepackage{graphicx}
\usepackage{mathptmx}      
\journalname{Few-Body Systems (APFB2011)}
\begin{document}

\title{\boldmath
Signature of strange dibaryon in kaon-induced reaction
}


\author{Shota Ohnishi         \and
        Yoichi Ikeda              \and
        Hiroyuki Kamano             \and
        Toru Sato
}


\institute{S. Ohnishi \and T. Sato \at
              Department of Physics, Osaka University, Osaka 560-0043, Japan \\
              \email{sonishi@kern.phys.sci.osaka-u.ac.jp}           
           \and
           Y. Ikeda \at
              Department of Physics, Tokyo Institute of Technology, Tokyo 152-8551, Japan
		   \and
		   H. Kamano \at
			Research Center for Nuclear Physics (RCNP), Osaka University, Osaka 567-0047, Japan
}

\date{Received: date / Accepted: date}

\maketitle

\begin{abstract}
We examine how the signature of 
the strange-dibaryon resonances in the $\bar{K}NN-\pi\Sigma N$ system shows up in the scattering 
amplitude on the physical real energy axis within the framework of Alt-Grassberger-Sandhas (AGS) equations. 
The so-called point method is applied to handle 
the three-body unitarity cut in the amplitudes. We also discuss the possibility that the strange-dibaryon 
production reactions can be used for discriminating between existing models of the two-body $\bar{K}N-\pi\Sigma$ 
system with $\Lambda(1405)$.
\end{abstract}

\section{Introduction}
\label{intro}
The structure of $\Lambda(1405)$ with spin-parity $J^\pi =1/2^-$ and strangeness $S=-1$ 
has been studied for a long time. 
In the constituent quark model, $\Lambda(1405)$ might be considered as p-wave excited state with $uds$ quarks. 
However, since the mass of $\Lambda(1405)$ is about 30 MeV below the $\bar{K}N$ threshold, 
it has also been suggested that $\Lambda(1405)$ is the s-wave $\bar{K}N$ quasi bound state due to 
the strongly attractive s-wave interaction of the $\bar{K}N$ system with $I=0$ \cite{Ref1}. 
Akaishi and Yamazaki suggested that this strong attraction will produce a new type of nuclei, 
the deeply bound kaonic nuclei \cite{Ref2}. 
The simplest deeply bound kaonic nuclei are 
the strange dibaryon, which are the resonances in the $\bar{K}NN-\pi \Sigma N$ system. 
The strange-dibaryons will give a baseline for the systematic investigation of such deeply 
bound kaonic nuclei, because the many body dynamics can be treated accurately. 
The strange dibaryon resonances have been studied with the Alt-Grassberger-Sandhas (AGS) 
equations \cite{Ref3,Ref4} and 
with the variational method \cite{Ref5,Ref6,Ref7} using the phenomenological meson-baryon 
interactions \cite{Ref3,Ref5,Ref6} or interactions based on the effective chiral 
Lagrangian \cite{Ref4,Ref7}. All the analyses suggest the existence of the strange-dibaryon resonances. 

The strange dibaryon resonances can be produced by photon- or kaon-induced reactions on light nuclei 
such as $d$ and $^3$He, 
and the signal of the resonances may be observed in the invariant mass and/or missing mass distributions of the decay 
products. Theoretical studies of the kaon-induced reactions have been done by Koike-Harada 
and Yamagata {\it et al.} within the
optical potential approach \cite{Ref8,Yamagata-2009}.

In this contribution, we present how the 
signature of the strange-dibaryon resonances in the $\bar{K}NN-\pi\Sigma N$ system shows up  in the three-body scattering 
amplitude obtained by solving AGS equations on the physical real energy axis, which is the 
basic ingredient to calculate the cross sections for strange-dibaryon production reactions measured in the experimental facilities 
such as J-PARC.

\section{Three-body Scattering Equations}
\label{sec:1}
\subsection{AGS equations}
\label{sec:2}
\begin{figure}
	\begin{minipage}{0.5\hsize}
		\begin{center}
			\includegraphics[width=0.8\textwidth,clip]{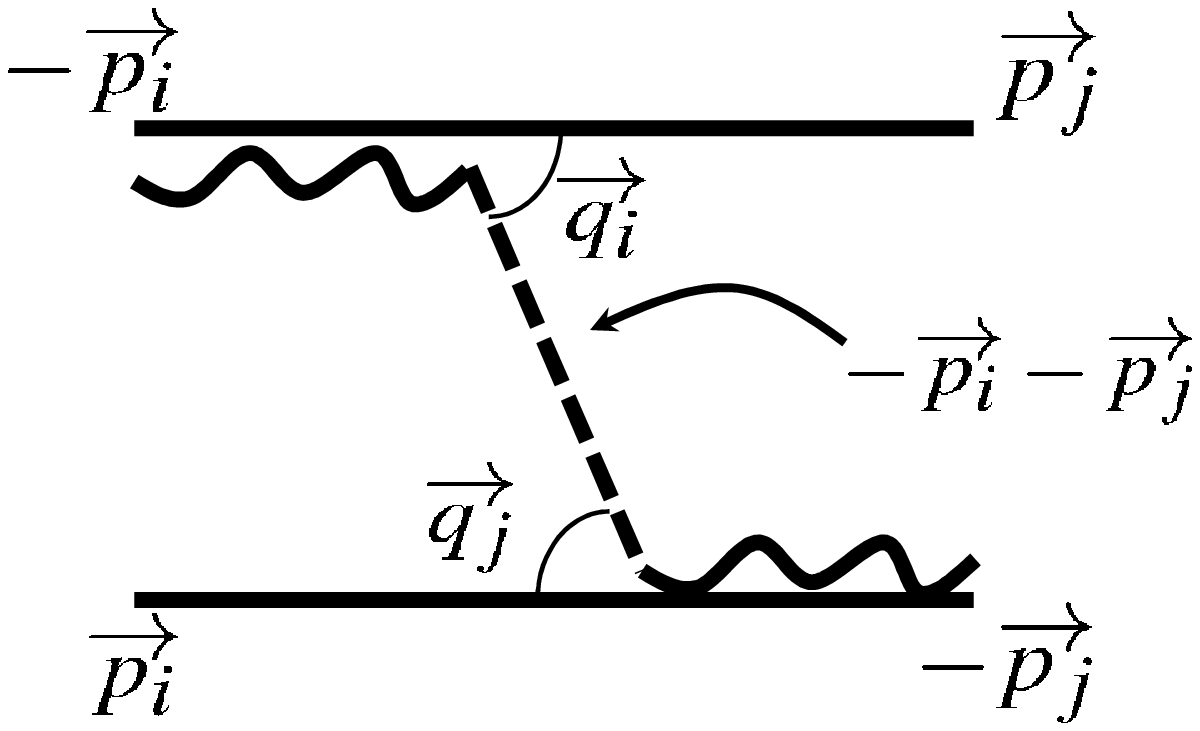}
		\end{center}
		\caption{One particle exchange interaction $Z_{i,j}({\bf p}_i,{\bf p}_j,W)$.}
		\label{fig:1}
	\end{minipage}
	\begin{minipage}{0.5\hsize}
		\begin{center}
			\includegraphics[width=0.6\textwidth,clip]{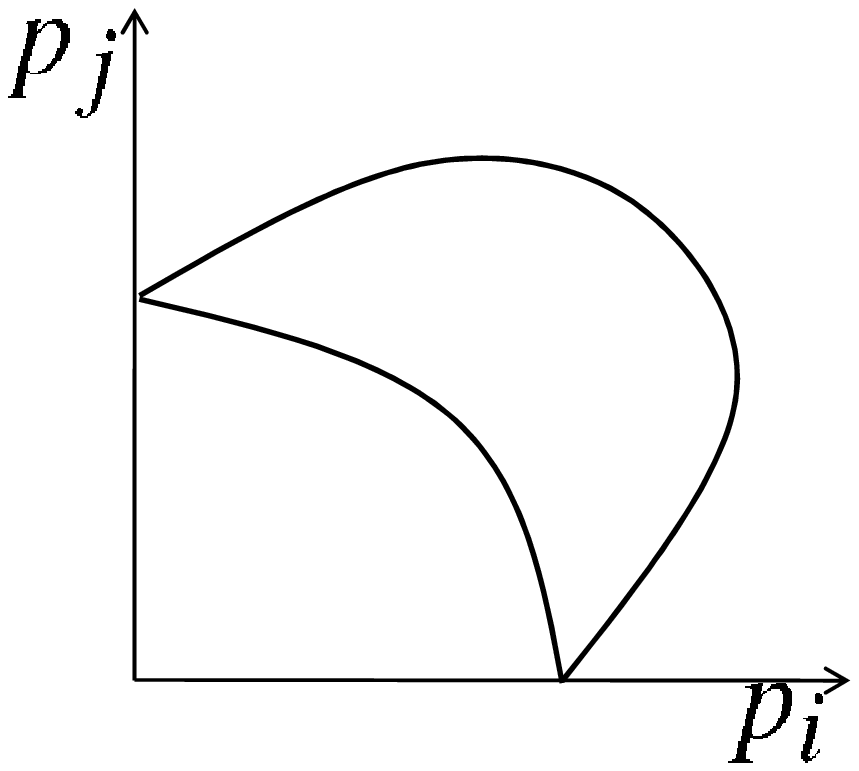}
		\caption{The moon-shape singularities. Solid curve shows the momenta $(p_i,p_j)$ 
		where $Z$ has logarithmic singularity.}
		\label{fig:2}
		\end{center}
	\end{minipage}
\end{figure}

The coupled channel equation for the $\bar{K}NN-\pi\Sigma N$ coupled channel system 
is given by the AGS equation
\begin{equation}
X_{i,j}({\bf p}_i,{\bf p}_j,W)=(1-\delta_{i,j})Z_{i,j}({\bf p}_i,{\bf p}_j,W)
+\sum_{n\ne i}\int d{\bf p}_nZ_{i,n}({\bf p}_i,{\bf p}_n,W)
\tau_n(W-E_n)X_{n,j}({\bf p}_n,{\bf p}_j,W),
\label{AGS}
\end{equation}
with the separable approximation for the interaction $V$
\begin{equation}
V({\bf q}',{\bf q})=\lambda g({\bf q}')g({\bf q}).
\end{equation}
Here $X_{i,j}({\bf p}_i,{\bf p}_j,W)$ is the quasi two-body amplitudes with the 
particle $i$ ($j$) as the spectator in the final (initial) state; the energy $W$ 
contains the infinitesimal positive imaginary part, 
$W=W'+i\epsilon$ with a real $W'$ and a infinitesimal positive $\epsilon$, 
resulting from the boundary condition of the scattering problem. 
The driving term $Z_{i,j}({\bf p}_i,{\bf p}_j,W)$ for the s-wave depicted in 
Fig.\ref{fig:1} is the particle-exchange interaction given by
\begin{equation}
Z_{i,j}(p_i,p_j,W)=2\pi \int^1_{-1} 
d(\cos \theta)\frac{g(q_i)g(q_j)}
{W-\frac{p_i^2}{2m_i}-\frac{p_j^2}{2m_j}-\frac{p_i^2+p_j^2+2p_ip_j\cos \theta}{2m_k}},
\end{equation}
where $\cos \theta=\hat{\bf p}_i\cdot \hat{\bf p}_j.$ 
Because of the three-body propagator in the integrand of this equation, the interaction 
$Z_{i,j}( p_i,p_j,W)$ has logarithmic singularities 
in $(p_i,p_j)$ plane known as the moon-shape singularities shown in Fig.\ref{fig:2}. 
Methods to handle these singularities are well studied, for example, 
with the spline method \cite{Ref9} or the point method \cite{Ref10,Ref11}. 
In this work, we employ the point method.
\subsection{Point method}
\label{sec:3}
The point method has been proposed by Schlessinger \cite{Ref10} and developed by 
Kamada \textit{et al.} \cite{Ref11}. 
We evaluate the amplitudes $X_{i,j}(p_i,p_j,W)$ in Eq.(\ref{AGS}) at 
$W=W'+i\epsilon_i$ with a real $W'$ and 
a finite positive $\epsilon_i$ ($i=1,2,...$). 
With finite $\epsilon_i$, the logarithmic singularities in $Z_{i,j}( p_i,p_j,W)$ 
become mild and numerical calculations can be performed safely. 
Then, we use the following continued fraction formula to extrapolate the amplitudes to the 
energy at $W=W'+i\epsilon$ with the infinitesimal positive $\epsilon$:
\begin{eqnarray}
X(W'+i\epsilon)&=&\frac{X(W'+i\epsilon_1)}{1+\frac{a_1(\epsilon-\epsilon_1)}{1+\cdots}}\\
&=&\frac{X(W'+i\epsilon_1)}{1+}\frac{a_1(\epsilon-\epsilon_1)}{1+}\frac{a_2(\epsilon-\epsilon_2)}{1+}\cdots,
\nonumber
\end{eqnarray}
with
\begin{eqnarray}
a_l=\frac{1}{\epsilon _l-\epsilon _{l+1}}\bigg( 1+
\frac{a_{l-1}(\epsilon _{l+l}-\epsilon _{l-1})}{1+}\frac{a_{l-2}(\epsilon _{l+1}-\epsilon _{l-1})}{1+}
\cdots
+\frac{a_1(\epsilon _{l+1}-\epsilon _1)}{1-[X(W'+i\epsilon _1) /X(W'+i\epsilon _{l+1})]}\bigg).
\end{eqnarray}

\section{Model of Two-body Interaction}
\label{sec:4}
We employ the two models for the meson-baryon interaction of the 
$\bar{K}NN-\pi\Sigma N$ system.
One is the model with the energy independent (E-indep.) separable potentials employed in \cite{Ref4}
\begin{eqnarray}
V_{\alpha \beta}(q',q)
=&-&\lambda_{\alpha\beta}\frac{1}{32\pi^2 F_\pi^2}
\frac{m_\alpha +m_\beta}{\sqrt{\mathstrut m_\alpha m_\beta}}
\left( \frac{\Lambda_\alpha^2}{q'^2+\Lambda_\alpha^2} \right)
\left( \frac{\Lambda_\beta^2}{q^2+\Lambda_\beta^2} \right),
\end{eqnarray}
and another is the model with the energy dependent (E-dep.) potentials employed in \cite{IKS-2010}
\begin{eqnarray}
V_{\alpha \beta}(q',q;E)
=&-&\lambda_{\alpha\beta}\frac{1}{32\pi^2 F_\pi^2}
\frac{2E-M_\alpha -M_\beta}{\sqrt{\mathstrut m_\alpha m_\beta}}
\left( \frac{\Lambda_\alpha^2}{q'^2+\Lambda_\alpha^2} \right)
\left( \frac{\Lambda_\beta^2}{q^2+\Lambda_\beta^2} \right).
\end{eqnarray}
Here, $\alpha$ and $\beta$ specify the meson-baryon channels, $m_\alpha$ ($M_\alpha$) is 
the meson (baryon) mass of the channel $\alpha$; $q$ and $q'$ are the relative momenta 
of the channels $\alpha$ and $\beta$ in the center of mass system, respectively; 
$F_\pi$ is the pion decay constant; 
$E$ is the total scattering energy of the meson-baryon system, which is determined by 
$W-p^2/2\eta$ with $\eta$ being the reduced mass between spectator particle and meson-baryon pair 
in the three-body system; 
$\lambda_{\alpha\beta}$ is determined by the 
flavor SU(3) structure of the Weinberg-Tomozawa term, assuming the different off-shell behavior 
with non-relativistic kinematics. 
\footnote{
In deriving the potentials from the Weinberg-Tomozawa term, 
we have also assumed $E_\alpha/M_\alpha \sim 1$ for baryons.}
Also, we have introduced the cutoff parameter $\Lambda_\alpha$. These parameters are determined 
by fitting the $\bar{K}N$ cross sections (The resulting values of the parameters are 
listed in Table\ref{tab:1}). 
\begin{table}[t]
\caption{Cutoff parameters of $\bar{K}N-\pi\Sigma$ interaction.}
\centering
\label{tab:1}       
\begin{tabular}{lllll}
\hline\noalign{\smallskip}
 &$\Lambda ^{I=0}_{\bar{K}N}$(MeV) &$\Lambda ^{I=0}_{\pi\Sigma}$(MeV) &$\Lambda ^{I=1}_{\bar{K}N}$(MeV) &$\Lambda ^{I=1}_{\pi\Sigma}$(MeV) \\[3pt]
\tableheadseprule\noalign{\smallskip}
E-indep.&1000 & 700 & 920&960 \\
E-dep.&1000 & 700 & 725&725 \\
\noalign{\smallskip}\hline
\end{tabular}
\end{table}
Here we take ``non-relativistic kinematics'' in this report.

\begin{figure}
	\begin{minipage}{0.5\hsize}
		\begin{center}
			\includegraphics[width=\textwidth,clip]{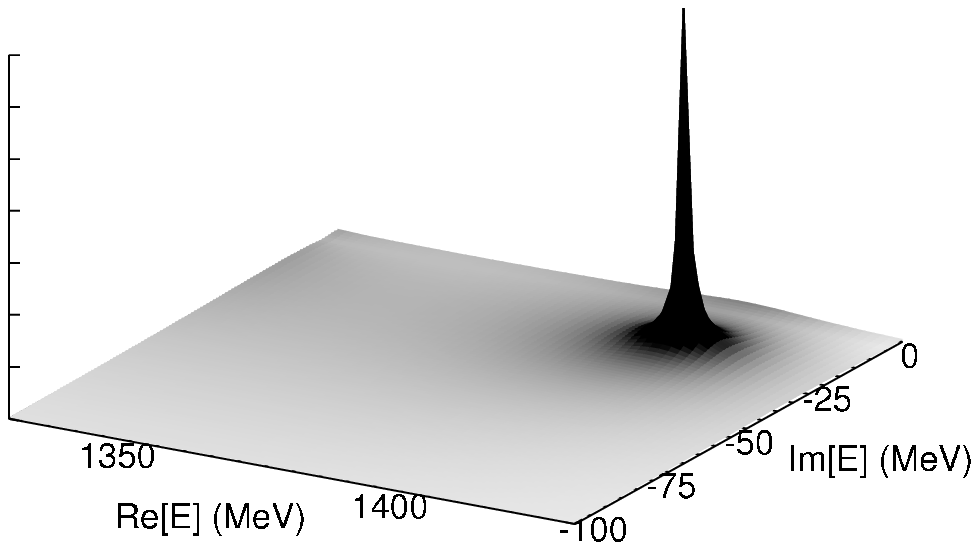}
			(a)
		\end{center}
	\end{minipage}
	\begin{minipage}{0.5\hsize}
		\begin{center}
			\includegraphics[width=\textwidth,clip]{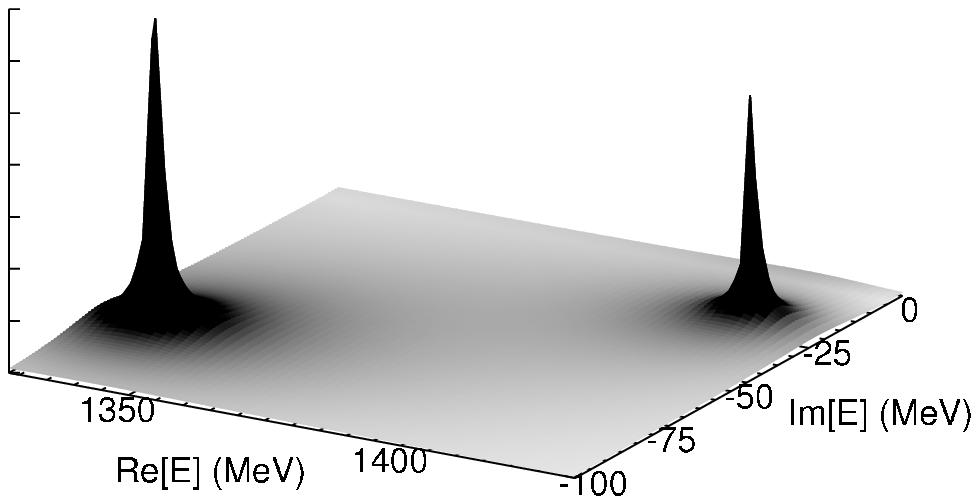}
			(b)
		\end{center}
	\end{minipage}
		\caption{The $S=-1$, $J^\pi =1/2^-$ $\bar{K}N$ s-wave amplitude on 
		complex energy plane in
		(a) the E-indep. model and (b) the E-dep. model.}
		\label{fig:3}
\end{figure}

We find that the above two models have a quite different analytic structure of the 
$\bar{K}N$ s-wave amplitude in the complex energy plane below the $\bar{K}N$ and above the 
$\pi\Sigma$ threshold energies:  
the E-indep. model has a pole 
corresponding to $\Lambda(1405)$ in the $\bar{K}N$ physical and 
$\pi\Sigma$ unphysical sheet (Fig.\ref{fig:3}(a)),
while the E-dep. model has two poles in the same sheet (Fig.\ref{fig:3}(b)). 
The analytic structure of the E-dep. model is similar to that of the chiral unitary model \cite{Ref12}. 
It will be then interesting to examine how this difference between the models of 
the two-body interaction emerges in the strange-dibaryon production reactions.

\section{Results and Discussion}
\label{sec:5}
\begin{figure}
	\begin{minipage}{0.5\hsize}
		\begin{center}
			\includegraphics[width=\textwidth,clip]{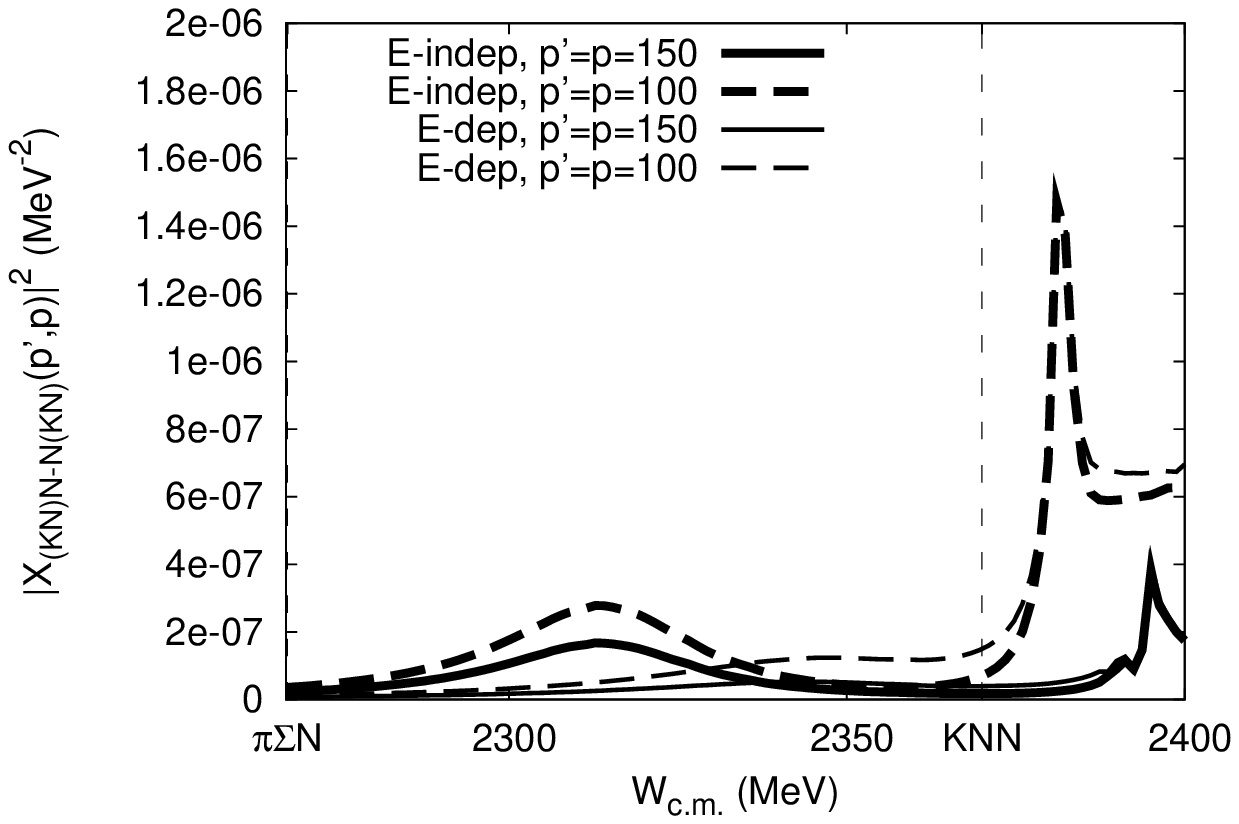}
			(a)
		\end{center}
	\end{minipage}
	\begin{minipage}{0.5\hsize}
		\begin{center}
			\includegraphics[width=\textwidth,clip]{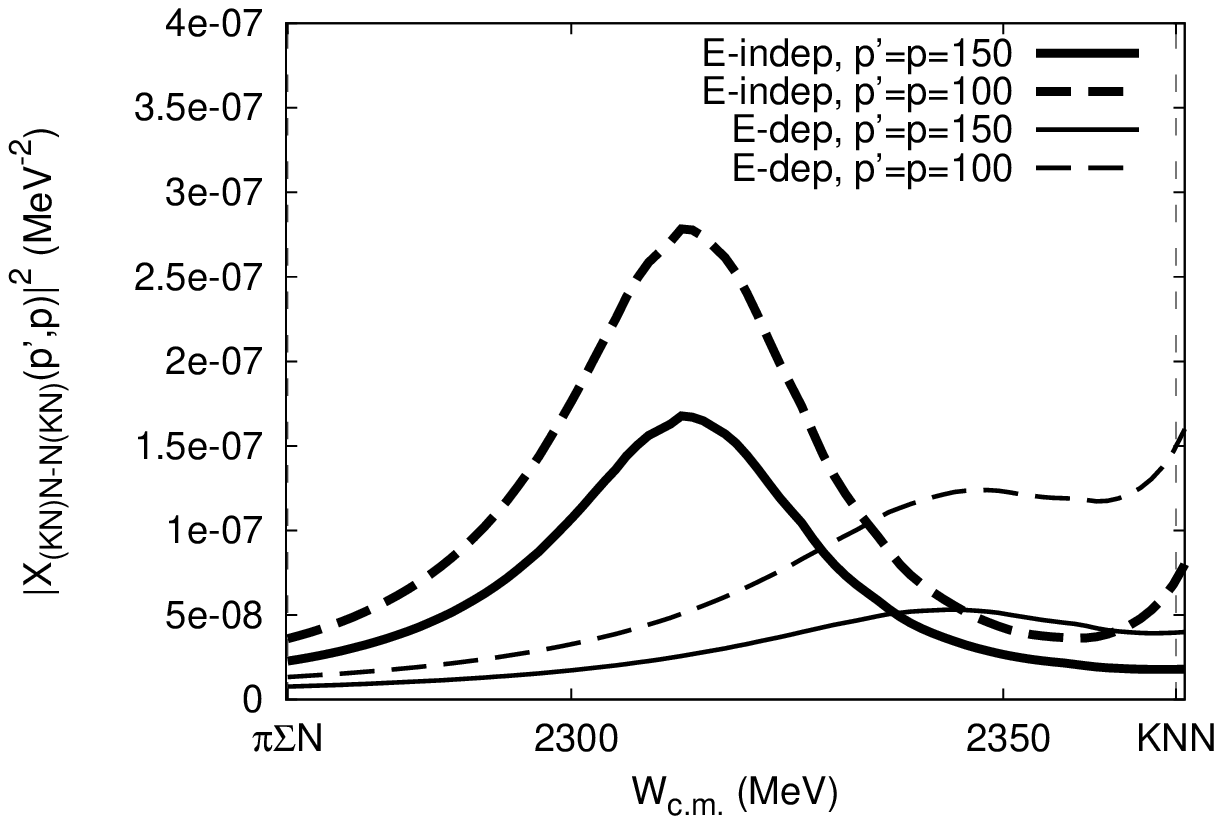}
			(b)
		\end{center}
	\end{minipage}
		\caption{Energy dependence of $|X_{(\bar{K}N)N-(\bar{K}N)N}|^2$.
				Dashed lines represent the amplitude at 100 MeV, and 
				solid lines represent the amplitude at 150 MeV. 
				Thick lines are E-indep. model, and thin lines are E-dep. model. 
				(b) is enlarged view of (a).}
		\label{fig:4}
\end{figure}

In this report, we presents the quasi two-body amplitudes by using the most important 
interactions. 
For three-body $Z$, we include 
$\bar{K}$-exchange mechanism but not $\pi$ or baryon exchange mechanism.
For two-body interaction, we include $\bar{K}N-\pi\Sigma$ interaction.
In Fig.\ref{fig:4}, we show $|X_{(\bar{K}N)N-(\bar{K}N)N}(p_i,p_j,W)|^2$ on the real 
energy axis for the E-indep. (thick curves) and E-dep. (thin curves). 
We observe a peak around $W\sim 2310$ MeV for the E-indep. model and a bump around 
$W\sim 2350$ MeV for the E-dep model. 
These peak and bump appear near the calculated resonance energy of the strange-dibaryons ($W_R=2329.5-i23.3$ MeV for the E-indep. model and $W_R=2352.0-i22.5$ MeV for 
the E-dep. model). 
This result suggests that the signal of the strange-dibaryons can emerge as a clear peak or 
a bump of the cross sections, which can be calculated from the amplitude-square $|X|^2$ on 
the real energy axis. The peak structure is pronounced in the E-indep. model, while in 
the E-dep. model it is rather small and may not be possible to separate from the background 
contributions. This difference of the three-body amplitudes due to the model dependence of the two-body subsystem suggests that the strange-dibaryon production reactions could provide also the 
useful information on the $\bar{K}N-\pi\Sigma$ system. 

In summary, by making use of the point method, we have calculated the quasi two-body amplitude 
$X_{i,j}(p_i,p_j,W)$ on the real energy axis. 
We then have found the bump structure in the amplitude in the energy region where the 
strange-dibaryons are expected to exist, implying that the signal of the strange-dibaryon 
resonances is possible to be observed in the physical cross sections. 
We have also shown that the strange-dibaryon production reactions could also be useful for 
judging existing dynamical models of $\bar{K}N-\pi\Sigma$ system with $\Lambda(1405)$. 
In the current work, however, we have not taken account of several reaction mechanisms such 
as $\pi$-exchanges. The further improvement of the current model and the calculation of the 
actual cross sections are under investigation.



\end{document}